# Spatiotemporally Localized Optical Links and Knots

Yaning Zhou[1,2,†], Nianjia Zhang[3,†], Ao Zhou[1,2], Zhao Zhang[1,2], Jinsong Liu[1,2], Chunhao Liang[1,2,*], Sergey A. Ponomarenko[4,5,*], Qiwen Zhan[3,6,*], Yangjian Cai[1,2,*], Xin Liu[1,2,*]

[1]Shandong Provincial Key Laboratory of Light Field Manipulation Physics and Applications & School of Physics and Electronics, Shandong Normal University, Jinan 250014, China.
[2]Collaborative Innovation Center of Light Manipulations and Applications, Shandong Normal University, Jinan 250358, China.
[3]School of Optical-Electrical and Computer Engineering, University of Shanghai for Science and Technology, Shanghai 200093, China.
[4]Department of Physics and Atmospheric Science, Dalhousie University, Halifax B3H 4R2, Canada.
[5]Department of Electrical and Computer Engineering, Dalhousie University, Halifax B3J 2X4, Canada.
[6]Zhejiang Key Laboratory of 3D Micro/Nano Fabrication and Characterization, Department of Electronic and Information Engineering, School of Engineering, Westlake University, Hangzhou 310030, China.
[†]These authors contributed equally to this work.
[*]Corresponding authors: chunhaoliang@sdnu.edu.cn; serpo@dal.ca; qwzhan@usst.edu.cn; yangjiancai@sdnu.edu.cn; xinliu@sdnu.edu.cn

**Optical links and knots have attracted growing attention owing to their exotic topologic features and promising applications in next-generation information transfer and storage. However, current protocols for optical topology realization rely on paraxial propagation of spatial modes, which inherently limits their three-dimensional topological structures to longitudinal space-filling. In this work we propose and experimentally demonstrate a scheme for creating optical knots and links that are localized in space within a transverse plane of a paraxial field, as well as in time. These spatiotemporal topological structures arise from polychromatic wave fields with tightly coupled spatial and temporal degrees of freedom that can be realized in the form of superpositions of toroidal light vortices of opposite topological charges. The (2+1)-dimensional nature of a toroidal light vortex imparts spatiotemporally localized wave fields with nontrivial topological textures, encompassing both individual and nested links or knots configurations. Moreover, the resulting topological textures are localized on an ultrashort timescale propagate at the group velocity of the wave packets and exhibit remarkable topological robustness during propagation as optical carriers. The nascent connection between spatiotemporally localized fields and topology offers exciting**

prospects for advancing space-time photonic topologies and exploring their potential applications in high-capacity informatics and communications.

## Introduction

In mathematical terms, linked and knotted topologies describe the ways in which closed curves can be arranged and intertwined in three-dimensional space or within a physical medium. Such nontrivial topological configurations are ubiquitous, emerging across a diverse array of physical systems, including classical fluid dynamics [1,2], liquid crystals [3-5], Bose-Einstein condensates [6,7], and various quantum fields [8-10]. Similar topological features also occur in optical fields, where they arise as three-dimensional trajectories of field lines [11-13], as well as phase and polarization singularities concentrated along zero-intensity regions [14-16]. Recent advances in light beam shaping have enabled the generation of polychromatic waves with knotted and linked topologies of optical fields [13,14,17-19], offering new avenues for robust information encoding [20,21]. Although knot theory has a long and venerable mathematical history [22], optical implementations of knots typically require paraxial propagation of a coherent superposition of two-dimensional transverse spatial modes [13-15,19-25]. The resulting topological structures are confined to a fixed three-dimensional spatial volume ($x - y - z$) rather than being localized. Thus, such structures remain statically embedded in space and hence cannot be transported along a communication channel. Consequently, the conventional optical topologies cannot serve as truly independent carriers, limiting their applicability to tasks such as topology-encoded optical transport [20,21].

Synthesizing spatiotemporally localized optical topologies inherently requires incorporating the temporal dimension, which would allow for the transport along a propagation axis. Recent impressive advances in hyper-dimensional light shaping have significantly enhanced our ability to tame light in both space and time [26,27]. Notable examples include space-time light sheets with nondiffracting properties [28-30], spatiotemporal optical vortices (STOVs) carrying transverse orbital angular momentum [31-33], toroidal vortices [34-36] and STOV pulse combs [37]. In this context, spatiotemporal optical wave packets, which are localized light fields in a (2+1)-dimensional space-time domain ($x - y - \tau$) with $\tau = t - z/v_g$ representing a local time in the reference frame co-propagating with the group velocity $v_g$, offer new perspectives for sculpting the topology of light fields [26]. Recently, numerous efforts have been made to imprint nontrivial topological textures onto ultrafast space-time fields, including skyrmions, merons and torons [38-45]. However, owing to the lack of a clear correspondence between mathematical and physical topology, identifying counterparts of nontrivial optical links and knots localized in both space and time remains elusive, and the stability of their propagation dynamics continues to pose significant challenges.

Here, we design a family of spatiotemporal optical links (STOLs) and knots (STOKs), localized in both space and time, and embedded within ultrashort, space-time nonseparable wave packets moving at the group velocity. We experimentally demonstrate that the complex fields of these spatially and temporally localized topologies can be physically realized as toroidal light vortices (TLVs) characterized by dual indices: poloidal and toroidal quantum numbers, thereby enabling the construction of a hierarchy of nested topological structures. Moreover, we further investigate the propagation dynamics of these wave packets, treated as individual optical carriers, in free space and linear dispersive media. The results reveal that these spatially and temporally localized topological waveforms exhibit remarkable robustness on propagation, maintaining their nontrivial topological textures over extended distances. This work breaks the stringent constraints of conventional optical knot theory, which relies on longitudinal beam propagation and concurrent transverse diffraction, marking a fundamental transition from topological photonic structures extended over the three-dimensional space to paraxial wave packets of nontrivial topology, physically localized in the transverse plane and time.

**Results**

**Concept of spatiotemporally localized optical topology**

A fundamental challenge for spatiotemporal optical topologies localized in space-time lies in transferring the parametrization of topology onto a localized framework, rather than trivially appending a temporal dimension [Methods]. To this end, we start by introducing a complex envelope of a space-time wave packet, localized in a three-dimensional space-time $(x, y, \tau = t - z/v_g)$ domain, which takes the form

$$\Psi(x, y, \tau) \propto \left(\sqrt{2}r_\perp/w_0\right)^{q_1} e^{-r_\perp^2/w_0^2} \cos(q_1\varphi_\perp + q_2\varphi_\parallel), \tag{1}$$

where $r_\parallel = \sqrt{x^2 + y^2}$ and $\varphi_\parallel = \tan^{-1}(y/x)$ are polar coordinates defined in the transverse plane; $r_\perp = \sqrt{(r_\parallel - r_0)^2 + \tau^2}$ and $\varphi_\perp = \tan^{-1}[\tau/(r_\parallel - r_0)]$ defined in the space-time region, and $q_1$ and $q_2$ quantify a twist number (i.e., topological charge) in the poloidal and toroidal planes, respectively. We can evaluate the intensity $S(x, y, \tau) \equiv |\Psi(x, y, \tau)|^2$ of this complex wave field as

$$S(r_\perp, \varphi_\perp, \varphi_\parallel) \equiv F(r_\perp) \cdot G_{q_1, q_2}(\varphi_\perp, \varphi_\parallel). \tag{2}$$

Here

$$F(r_\perp) = \left(\sqrt{2}r_\perp/w_0\right)^{2q_1} \exp(-2r_\perp^2/w_0^2), \tag{3}$$

is a radial function that describes a ring-shaped torus (see Fig. 1a) and

$$G_{q_1,q_2}(\varphi_\perp, \varphi_\parallel) = [1 + \cos(2q_1\varphi_\perp + 2q_2\varphi_\parallel)], \tag{4}$$

is an azimuthal function that gives rise to a series of three-dimensionally twisted surfaces that radially expand toward infinity (see Fig. 1b). The iso-intensity of the product of Eq. (3) and Eq. (4) forms a topological link or knot confined within a finite spatiotemporal torus (see Fig. 1**c**), with its structure determined by the topological charges, $q_1$ and $q_2$. Note that this configuration shares certain mathematical similarities with the corresponding structures in topology [Methods], but the latter typically defines topological features as zero lines of a complex polynomial in the spatial domain, making it challenging to directly extend such definitions to localized frameworks. The intensity in the toroidal plane must be single-valued, implying that $S(r_\perp, \varphi_\perp, \varphi_\parallel) = S(r_\perp, \varphi_\perp, \varphi_\parallel + 2\pi)$ which imposes a quantization rule, forcing $q_2$ to be either an integer or a half-integer. We can loosely speak of a transverse angular momentum of the system, quantified by $q_1$, as "orbital" and the longitudinal angular momentum, quantified by $q_2$, as "spin" because the latter describes rotation of the torus around its axis (the axis orthogonal to the toroidal plane). Interestingly, our "spin" must be either integer or half integer in (admittedly) a somewhat superficial analogy with the spin of a quantum particle.

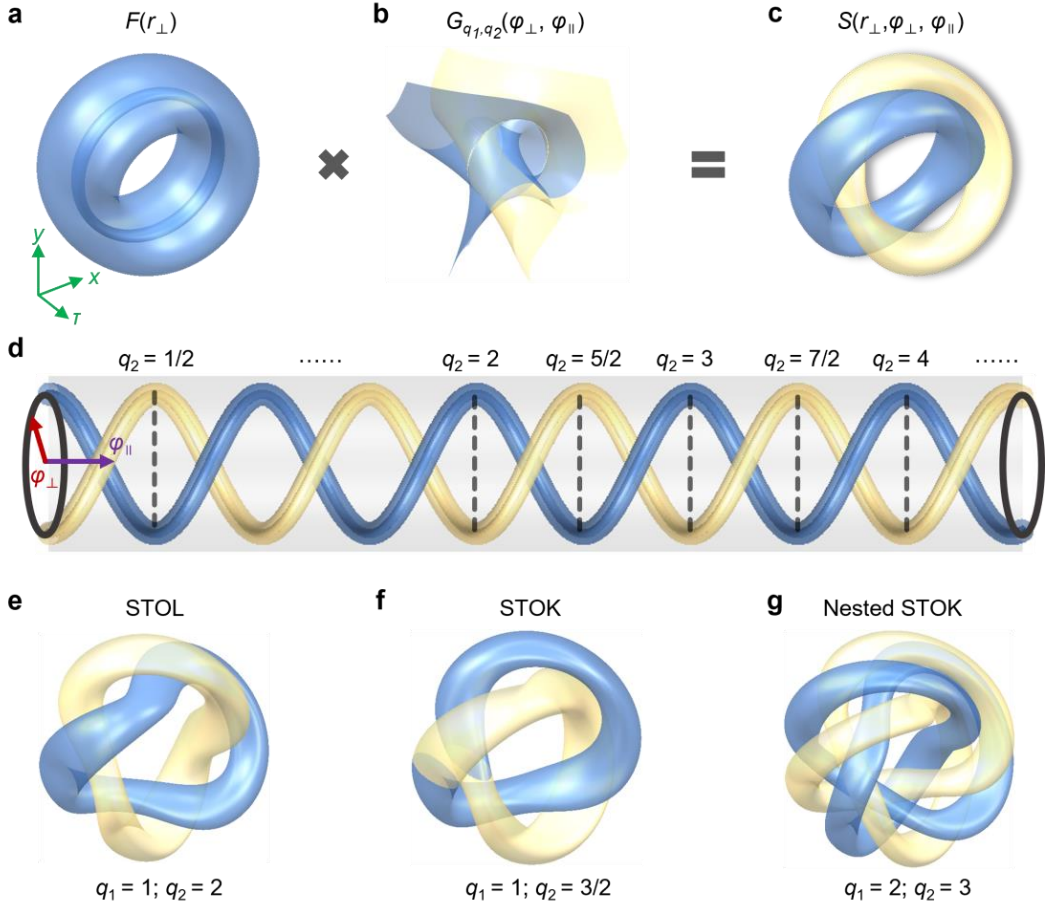

**Fig. 1 | Concept of optical topologies localized in space and time. a-c** A combination of a torus and a set of infinite twisted surfaces yields a topological structure embedded within a spatiotemporal iso-intensity surface. **d** Illustration of a spatiotemporal iso-intensity topology represented as a strand-periodic braid enclosed in a cylindrical volume, the phase difference of distinct colors being $\pi$. **e-g** Fundamental STOLs appear for integer $q_2$, fundamental STOKs appear for half-integer $q_2$, nested structures appear for integer multiples of $q_1$ and $q_2$. Strands of different colors are woven into two separate loops in the STOLs, whereas they form a single loop in the STOKs.

In Cartesian coordinates $(\varphi_\perp, \varphi_\parallel)$, the closed torus is unfolded into a cylinder, and the link or knot is mapped into an embedded braid comprised of twined strands (marked by blue and yellow), as illustrated in Fig. 1d. Thus, when $q_2$ is integer, the endpoints of each braided zero lie on the same strand, leading to a STOL structure shown in Fig. 1e. If $q_2$ is a half-integer, the endpoints lie on different strands, resulting in a STOK structure shown in Fig. 1f. We can infer from the figure that in this case, the strands within a STOL are linked to each other with a $\pi$ phase difference, whereas in STOKs, strands possessing a $\pi$ phase difference are connected, yielding a closed knot.

It follows, building on the above theory, that nested STOLs and STOKs [21] can also be constructed by scaling $Nq_1$ and $Nq_2$ by integer multiples, as shown in Fig. 1g and the corresponding braid representation in Supplementary Note 1.

We can identify the designed complex fields with physically realizable paraxial, narrow-band optical pulsed beams. Specifically, we can implement the wavefield giving rise to spatiotemporally localized optical topologies of Eq. (2) with a pair of conjugated TLVs of opposite "orbital" and "spin" angular momenta. In particular, $S \propto |\Psi_{\text{TLV}} + c.c.|^2$ where the field profile of each TLV is given by [35,47]

$$\Psi_{\text{TLV}}(x, y, \tau) = C_{q_1,q_2}\left(\sqrt{2}r_\perp/w_0\right)^{q_1} e^{-r_\perp^2/w_0^2} e^{iq_1\varphi_\perp} e^{iq_2\varphi_\parallel}, \tag{5}$$

where $\tau = t - z/v_g$ is a local time in the co-propagating frame with $v_g$; $C_{q_1,q_2}$ is a normalization constant, and $w_0$ is the rms beam width in the poloidal plane. Evidently, the wave field localized at a specific azimuthal angle $\varphi_\parallel$ forms a regular STOV wave packet with the off-axis topological charge $q_1$ relative to the toroidal vortex line radius $r_0$. The number $q_2$ quantifies the topological charge in the toroidal plane.

It is worth noting that the topologies of Eq. (2) are formed by three-dimensional iso-intensity surfaces with edge dislocations, which are fundamentally distinct from conventional optical topologies that are usually defined by field singularities corresponding to sets of zero-intensity points or lines within a complex field. Furthermore, while conventional optical knots and links are established via the evolution of monochromatic beams over extended distances in free space and constructed with the help of Milnor polynomials [14], the spatiotemporal intensity topologies that we introduce in this work are localized pulsed beam entities, sculpted in the space-time domain and confined to ultrashort time scales.

**Experimental generation of TVL and topology reconstruction**

Our theoretical analysis indicates that the preparation of TLV wave packets is the key to the experimental synthesis of STOLs and STOKs. These toroidal vortices form a class of three-dimensional, nonseparable in space and time, structured light fields, which are inherently challenging to generate directly. In our experiment, we first use a two-dimensional pulse shaper to produce a poloidal STOV pulse. This pulse is then transformed into a ring-shaped structure through a conformal mapping system (Supplementary Note 2), ultimately enabling the synthesis of TLVs, as illustrated in Fig. 2. A pulsed beam, centered at 1030 nm with a ~15 nm bandwidth, is shaped by a pulse shaper comprising a grating, cylindrical lens, and helical phase of $q_1$ encoded into SLM1. After modulation and recombination, a near-field poloidal STOV pulse is generated. This STOV is then axially stretched using an afocal cylindrical beam expander (magnification 1:7), forming an extended vortex tube. The STOV tube is subsequently converted into a closed-loop toroidal vortex via free-space propagation shaped by SLM2. A third modulator (SLM3) imparts an additional helical phase of $q_2$ and ensures field collimation. The resulting TLV is relayed to the camera plane using a 4f imaging system consisting of lenses L1 and L2.

To reconstruct the complete information of the object wave packet, a Fourier-transform-limited probe pulse—generated by the grating pair—is used to interfere with the object pulse via time-sliced off-axis interferometry [26]. By scanning the entire object wave packet, both the amplitude and phase distributions can be retrieved from a series of time-resolved interference fringes (Supplementary Note 3). Unfortunately, irregular fringe shifts exist between adjacent temporal slices. These shifts arise from inherent instabilities in the interferometer, primarily due to the motorized stage and minor optical misalignments, leading to random phase fluctuations across different time delays. To avoid this, we assume that the temporal phase at a specific spatial position along the $\tau$-axis remains constant. Accordingly, we normalize each sliced phase profile with respect to a particular spatial position. Supplementary Note 3 provides a detailed diagram of the three-dimensional reconstruction of spatiotemporal topologies. Finally, by stitching together the superpositions of all retrieved time slices and their corresponding complex conjugates, the three-dimensional spatiotemporal topology can be reliably reconstructed.

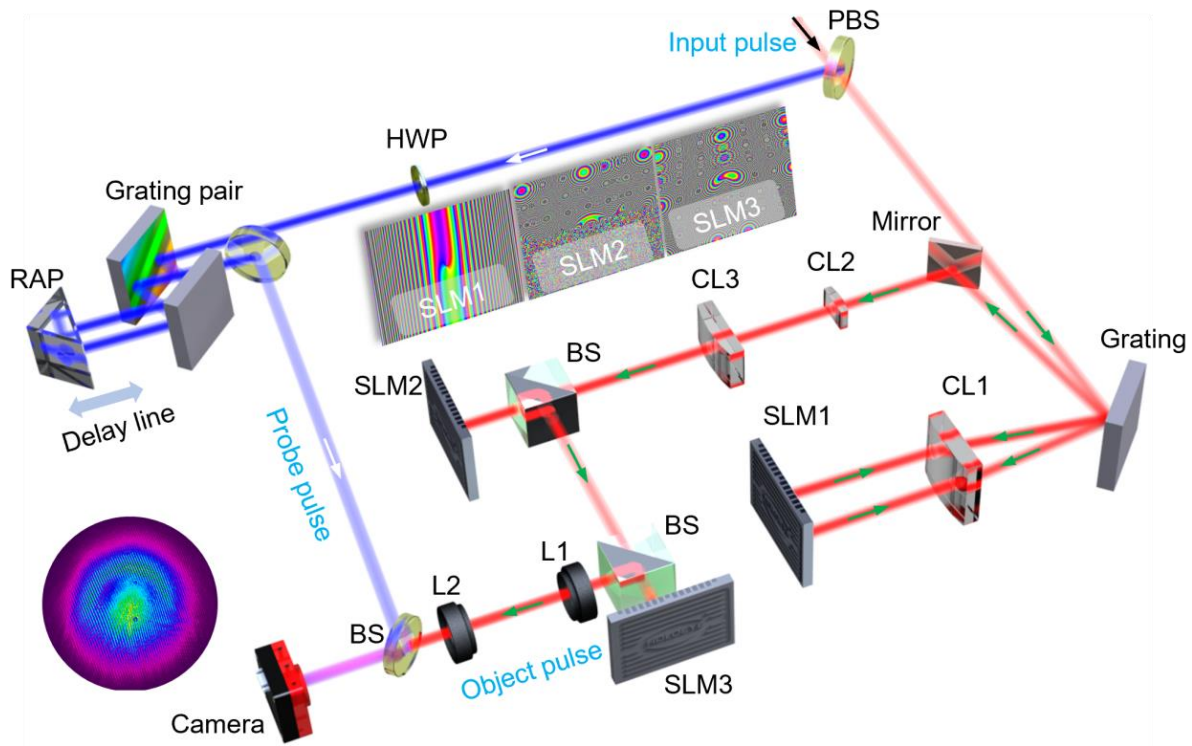

**Fig. 2 | Experimental setup for generating and characterizing STOLs and STOKs.** PBS: polarization beam splitter, HWP: half-wave plate, CL1-3: cylindrical lens, SLM1-3: spatial light modulator, BS: beam splitter, RAP: right-angle prism, L1-2: 4f optical imaging system.

**Theoretical and experimental results**

We first apply our model to construct STOLs theoretically and realize them experimentally. In Fig. 3, we exhibit an example corresponding to $q_1 = 1$ and integer $q_2$. The theoretical results, presented in Figs. 3a and 3b, follow from Eqs. (1)-(4). We obtain the corresponding experimental results, presented in Figs. 3c and 3d, via a synthesis of pulse shaping and confocal mapping; details of the experimental procedure are provided in Supplementary Note 3. We can conclude from the figures that the iso-intensity surfaces of the spatiotemporal wave packets form two independent, well-defined closed-loop tori localized in space and time. These two tori are mutually linked with a winding number of $q_2$ and exhibit a $\pi$ phase difference, known as an edge dislocation in singularity optics, as is indicated by the different colors in Figs. 3a and 3b. Although such topological configurations are visualized at specific iso-intensity levels (8%), the linked ring tori remain disjoint for any infinitesimal iso-intensity values, owing to the existence of a phase edge dislocation that enforces the braid separation (Supplementary Note 4). The experimental results agree well with the theoretical predictions.

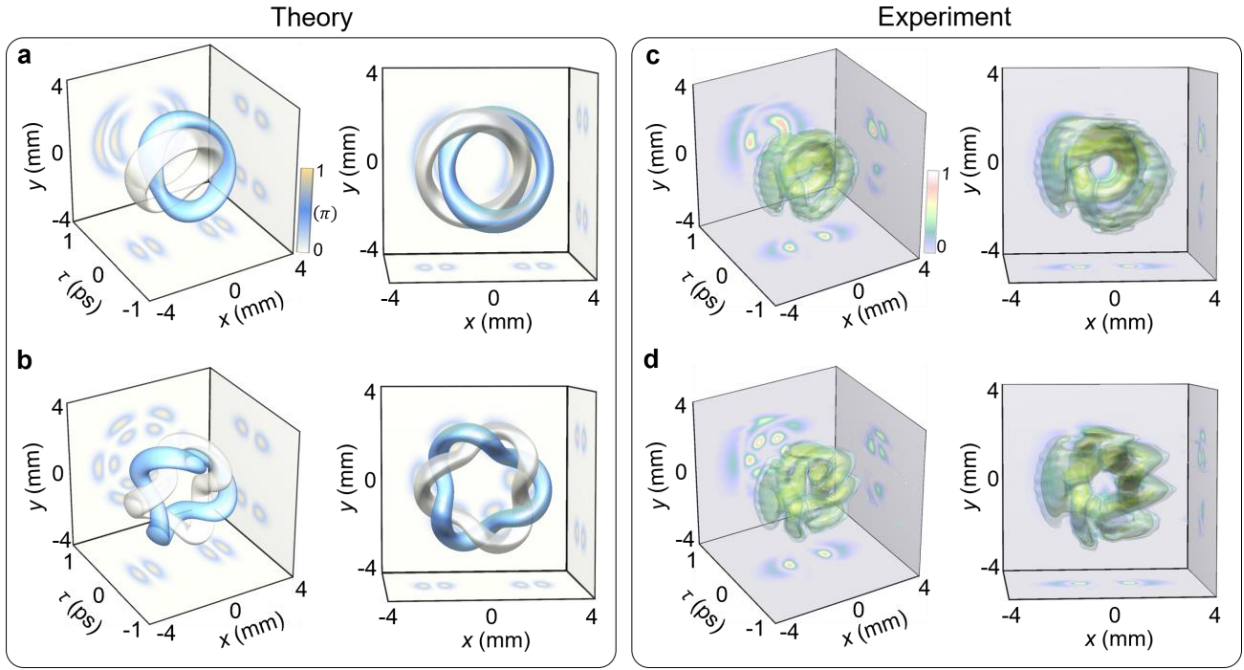

**Fig. 3 | Theoretical and experimental results of STOLs.** Three-dimensional iso-intensity surfaces of (**a** and **b**) theoretically designed and (**c** and **d**) experimentally produced space-time linked topologies from different viewpoints. **a** and **c** correspond to $q_1 = 1, q_2 = 1$; **b** and **d** to

$q_1 = 1, q_2 = 3$. The iso-intensity thresholds are set to 8%. The inset 2D intensity projections are taken at the cross-sections $x = 0, y = 0$ and $\tau = 0$, respectively.

If $q_2$ is chosen to be a half-integer, a knotted in space-time light field can also be constructed. In Fig. 4, we display both theoretical and experimental results for STOKs with half-integers $q_2$, showing representative examples with (Figs. 4a and 4c) $q_2 = 3/2$ and (Figs. 4b and 4d) $q_2 = 5/2$. We can infer from the figure that the resulting iso-intensity surface constitutes a single closed optical knot, formed by the junction of two segments that are $\pi$ out of phase with one another (Figs. 4a and 4b). These two segments are mutually twisted into each other $(2q_2 - 1)/2$ times, giving rise to a knot-like topology. All theoretically and experimentally generated STOLs and STOKs are consistent with those shown in Fig. 1.

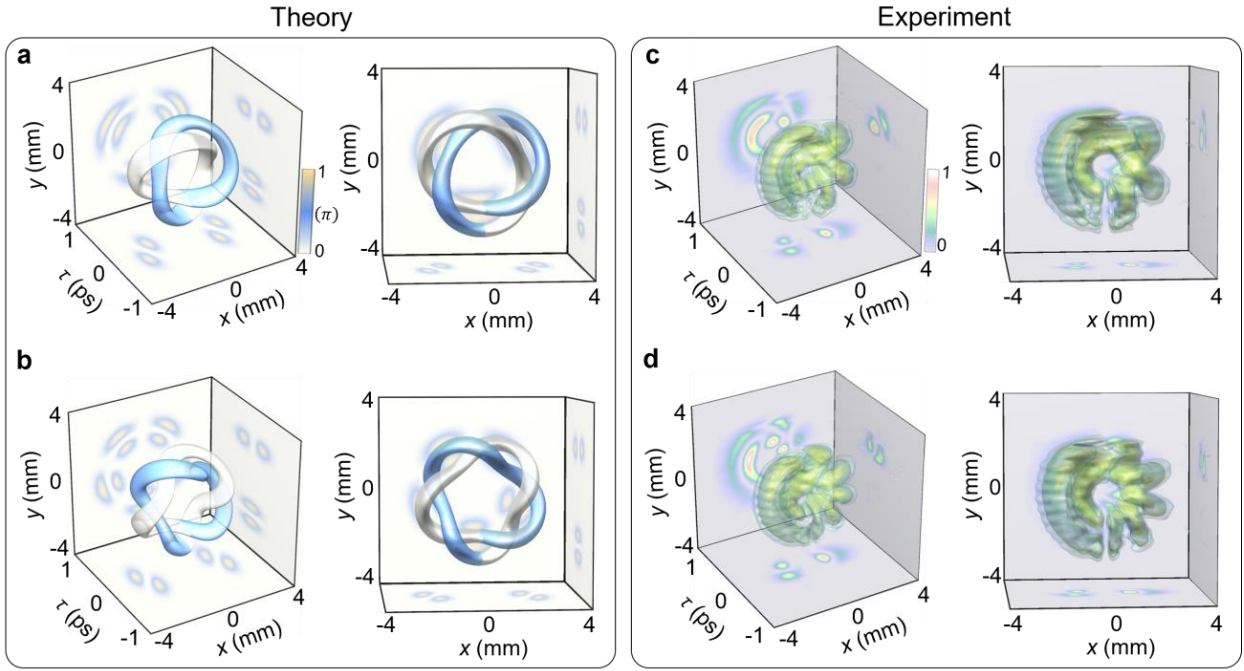

**Fig. 4 | Theoretical and experimental results of STOKs.** Three-dimensional iso-intensity surfaces of (**a** and **b**) theoretically and (**c** and **d**) experimentally constructed full space-time knotted topologies in different views. **a** and **c** Correspond to $q_1 = 1, q_2 = 3/2$; **b** and **d** Correspond to $q_1 = 1, q_2 = 5/2$. The iso-intensity thresholds are set to 8%. The inset 2D intensity projections are taken at cross-sections $x = 0, y = 0$ and $\tau = 0$, respectively.

Optical links and knots generated to date, have been largely limited to simple individual structures, primarily due to the lack of effective control over additional degrees of freedom of light [21]. Instructively, TLVs possess dual topological charges, offering new avenues for constructing nested and more intricate topological configurations. To demonstrate this capability, we show in Figure 5 theoretically constructed nested STOLs and STOKs, corresponding to higher-order (doubled) topological charges, $(2q_1, 2q_2)$, of the individual topological structures shown in Figs. 3 and 4, respectively. These nested, localized space-time optical topologies exhibit a distinct layer-by-layer hierarchical architecture whereby each layer consists of individual STOLs or STOKs. Regardless of whether a given layer is composed of fundamental STOLs or STOKs, the layers are interconnected in a linked configuration that imposes an identical phase distribution onto every strand. To provide a clearer illustration, we provide a detailed braid representation of Figure 5 in Supplementary Note 1.

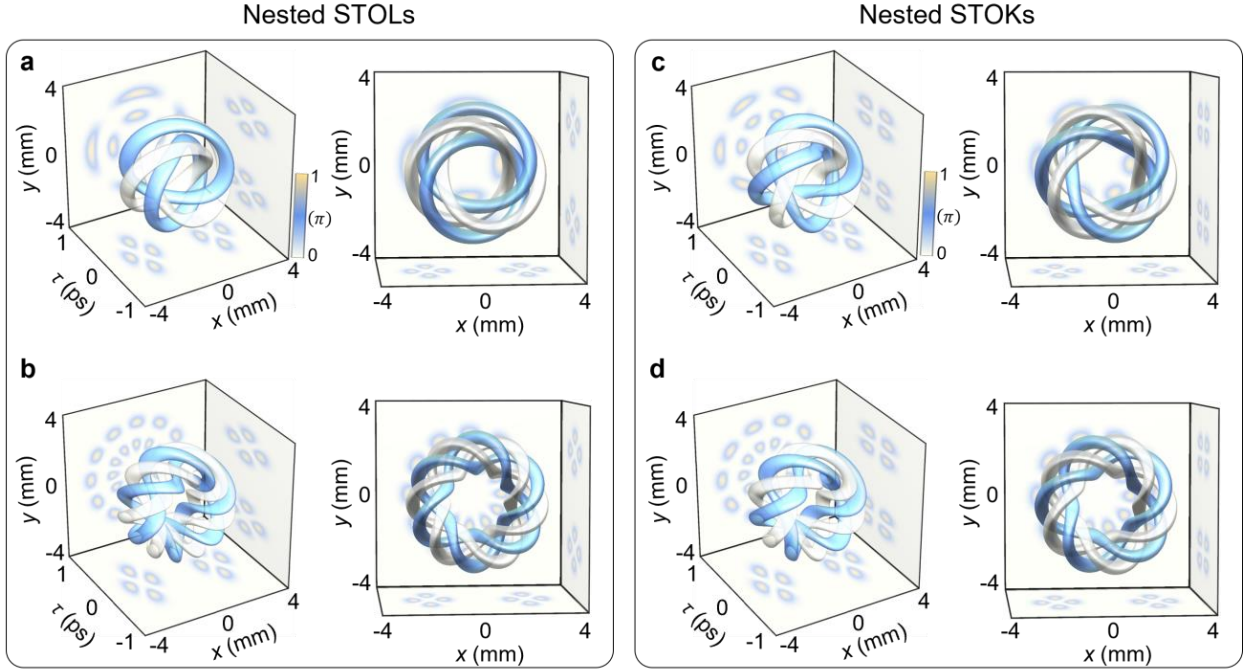

**Fig. 5 | Diagram of nested STOLs and STOKs. a** and **b** Correspond to STOLs with $q_1 = 2$, $q_2 = 2$ (upper) and $q_2 = 6$ (lower). **c** and **d** Correspond to STOKs with $q_1 = 2$, $q_2 = 3$ (upper) and $q_2 = 5$ (lower). Other parameters are identical to those used in Figs. 3 and 4.

Finally, we examine the evolution of an experimentally generated STOL with $(q_1 = 1, q_2 = 1)$ in free space and linear dispersive media using the angular spectrum method (Supplementary Note 5). Unlike conventional optical topologies that require long-distance propagation to form three-dimensional trajectories, spatiotemporally localized optical topologies are embedded into individual pulses that act as individual information carriers during transmission through dispersive media. In our simulations, we take silica glass as a typical propagation medium, with the corresponding group velocity dispersion (GVD) coefficients at different wavelengths 1550nm and 850nm taken as $\beta_2 = -28 \text{ fs}^2/\text{mm}$ and $\beta_2 = +32 \text{ fs}^2/\text{mm}$, respectively [46]. In Fig. 6, we exhibit the propagation dynamics of experimentally constructed STOLs in different dispersion regimes. We can infer from the figure that the STOL maintains a well-preserved linked structure under all propagation conditions. These results highlight the structural robustness and stability of optical topologies localized in space and time to propagation over long distances. This stability is underpinned by the fact that a vortex ring is an (approximate) self-similar solution to Maxwell's equations in the anomalous dispersion regime (Supplementary Note 6). Although the individual TLV is short lived [35,47] in free space or a linear medium with normal dispersion, our results reveal that superpositions of TLVs with opposite topological charges, which showcase a nontrivial topology, are well preserved ensuring overall morphological stability (Supplementary Note 6). In the normal dispersion case, in particular, the topological structure undergoes temporal inversion, which arises from the interplay between transverse and longitudinal OAM, leading to a reversal of the poloidal phase singularity [47].

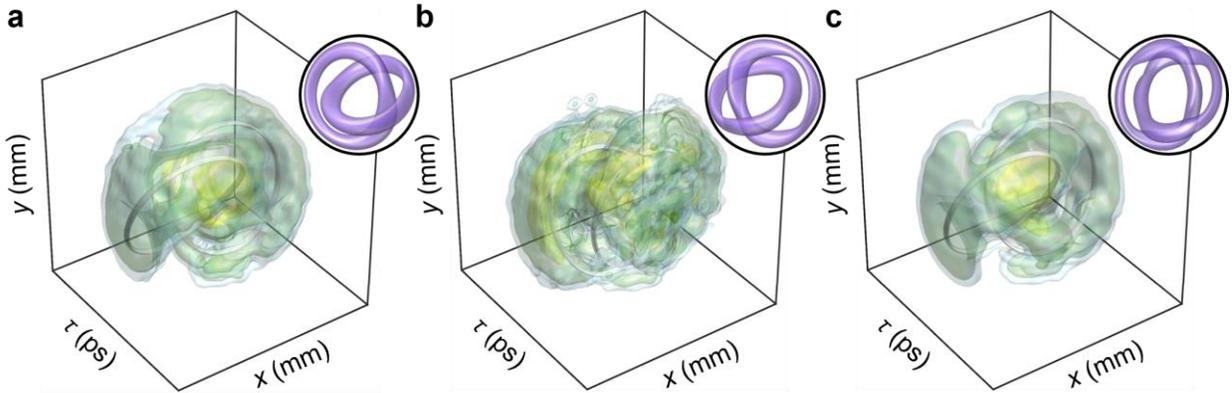

**Fig. 6 | Iso-intensity profile of an experimentally generated STOL propagating in distinct dispersion regimes and in vacuum at a distance of $2z_R$ from the source. a** STOL centered @1550 nm propagating in silica glass with GVD $\beta_2 = -28$ fs$^2$/mm. **b** STOL centered @850 nm propagating in silica glass with GVD $\beta_2 = +32$ fs$^2$/mm. **c** STOL centered @1550nm propagating in vacuum. The spatial and temporal beam widths are 0.95 mm and 350 fs (Supplementary Note 7), respectively. Iso-intensity surfaces are plotted at 4% of the peak intensity. The insets show the corresponding simulation results.

## Discussion

We have proposed and experimentally demonstrated a model for a hierarchy of spatiotemporal optical links and knots (STOLs and STOKs) that are localized on ultrashort time scales. These nontrivial topological textures are woven by the iso-intensity values of polychromatic spatiotemporal light fields and are experimentally realized using TLV pulses. Both theoretical and experimental results show that the topological structure of these spatiotemporally localized iso-intensity trajectories can be adjusted by tuning the poloidal and toroidal topological charges of the TLV, enabling the formation of both fundamental and nested STOLs and STOKs. Furthermore, we have shown robust propagation of these spatiotemporal topologies in free space and linear media in anomalous and normal dispersion regimes. Our findings offer a new paradigm for topological photonics in the spatiotemporal domain and may open up novel avenues for high-capacity information encoding, storage, and encryption in diverse platforms such as quantum optics, nonlinear optics, and condensed matter physics.

In contrast with the previous work, focused on spatial topological links and knots resulting from diffraction of monochromatic Laguerre-Gaussian modes [19-25], our approach unfolds in both space and time. Specifically, the spatiotemporal localization of optical topologies occurs through judicious engineering of space-time coupling of polychromatic wave fields, with no reliance on diffraction. In addition, we sculpt spatiotemporally localized topologies by tailoring the iso-intensity distribution along three-dimensional trajectories of the wave packet. In this approach, the topological paths are defined by peak intensity rather than null intensity, thereby enabling the generation of extremely intense fields [50] and the facilitation of nonlinear interactions [51]. Owing to its ultrashort timescale, this strategy also provides a brand-new means of potentially inscribing topological defects in photosensitive materials [52,53] using high-power femtosecond laser technology. Our work marks a significant advance in topological photonics because the realization of localized in space and time optical knots and links provides a robust and versatile platform for high-density data storage, encryption, and transport.

## Methods

**Revisiting conventional spatial and space-time localized topology from geometry analogy.**
Brauner [48] constructed an explicit class of complex scalar knotted fields corresponding to each $(m, n)$ torus knot or link, represented as a set of zeros of the complex polynomial $p(u, v) = u^m \pm v^n$, where $u$ and $v$ are complex variables, given by [49]

$$u = \frac{r^2 - 1 + 2iz}{r^2 + 1}, v = \frac{2(x + iy)}{r^2 + 1}, \tag{6}$$

where $(x, y, z)$ denote Cartesian coordinates and $r = \sqrt{x^2 + y^2 + z^2}$. We rewrite Eq. (6) as $u = |u|e^{i\varphi_\perp}$ with $\varphi_\perp = \tan^{-1}[2z/(r^2 - 1)]$ and $v = |v|e^{-i\varphi_\parallel}$ where $\varphi_\parallel = -\tan^{-1}(y/x)$ in cylindrical coordinates. To solve $|p(u, v)|^2 = 0$ and following Ref. [12], a torus knot or link arises from the condition $u^m \pm v^n = 0$, which implies $|u|^m = |v|^n \Rightarrow |u|^m - |v|^n = 0$. Under this condition, following Eq. (1), we have $|u| \neq 0$ and $|v| \neq 0$, and the necessary and sufficient condition for the topological curve described by the zeros of Eq. (1) is $|u|^m - |v|^n = 0$ and $|p(u, v)|^2 = 2|u|^m|v|^n[1 \pm \cos(m\varphi_\perp + n\varphi_\parallel)] = 0$. To map onto the local time frame, we combine the two conditions and define a new parameter function given by

$$S(x, y, z) = F(x, y, z)G_{m,n}(x, y, z), \tag{7}$$

with

$$F(x,y,z) = 2|u|^m|v|^n(|u|^m - |v|^n), \tag{8}$$

and

$$G_{m,n}(x,y,z) = [1 \pm \cos(m\varphi_\perp + n\varphi_\parallel)]. \tag{9}$$

It is worth noting that the zero trajectory of normalized $S(x,y,z)$ gives rise to a spatial topology [Fig. 7a] that has the same form as that of $p(u,v)$ [Fig. 7b]. The function $F(x,y,z)$ defines a toroidal surface analogous to that in the model described by Eq. (3), as we illustrate in Figs. 7c1 and 7c2. As $F(x,y,z)$ attains values arbitrarily close to zero, it yields a pair of distinct, closed iso-surfaces. At the zero-level set $F(x,y,z) = 0$, these two iso-surfaces coalesce, forming a unified three-dimensional torus. The function $G_{m,n}(x,y,z)$ describes a family of twisted surfaces that extend toward infinity in complete analogy with Eq. (4), as shown in Figs. 7d1 and 7d2. The product of these functions, given by $S(x,y,z) = F(x,y,z) \times G_{m,n}(x,y,z)$, results in a topological surface, illustrated in Fig. 7e1. This surface structure further reduces to a topological curve when $S(x,y,z) = 0$. Comparing Eqs. (2)-(4) and Eqs. (7)-(9), we observe that they have a similar mathematical structure with $m = 2q_1$ and $n = 2q_2$, as illustrated in Fig. 7e. A fundamental distinction exists between the two topological models: conventional spatial topologies are defined by zero-value trajectories in the purely spatial $(x,y,z)$ domain, where the structures are not genuinely localized in time but requires longitudinal beam propagation [13-15,19-25]. In contrast, spatiotemporally localized topologies are characterized by iso-intensity trajectories in the localized space-time domain $(x,y,t-z/v_g)$, as illustrated in Figs. 7e1 and 7e2. Overall, we derived the mathematical form of the spatiotemporal localized topological structure from the intrinsic similarity in the geometric representation of topological structures, rather than by trivially appending a time dimension, which would merely stretch the spatial topology along time.

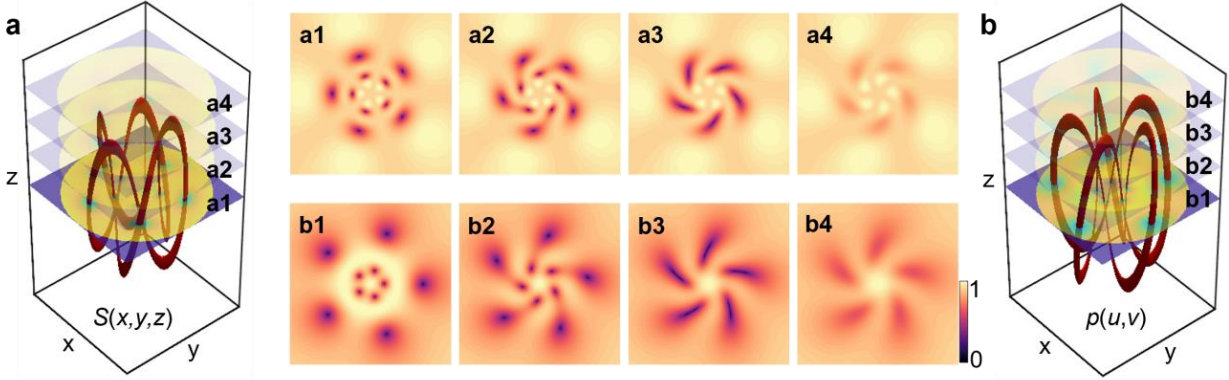
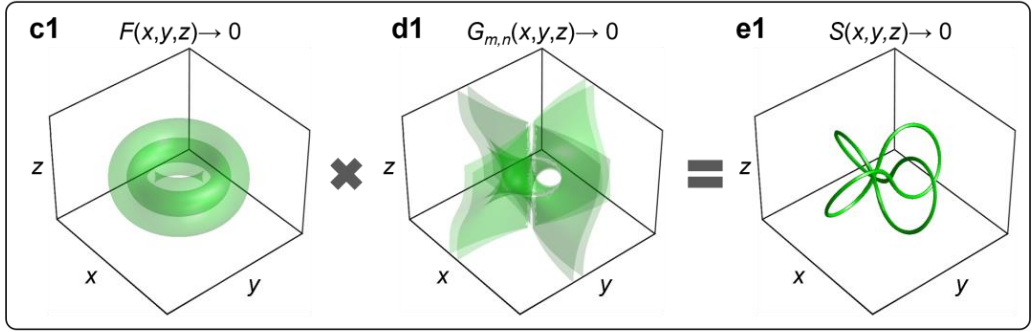
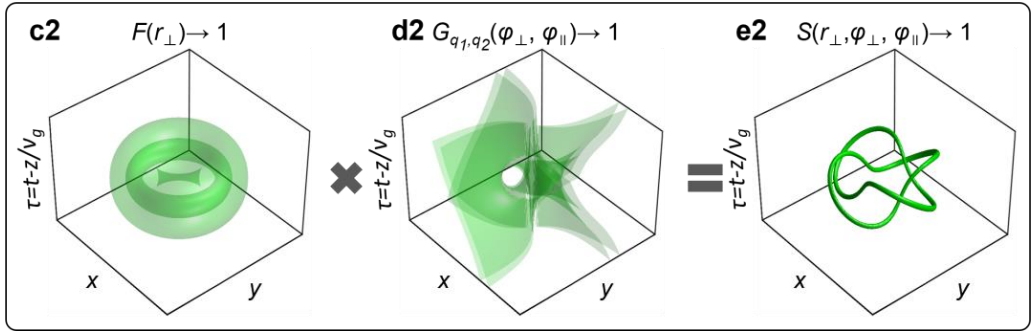

**Fig. 7 | Construction principle of conventional spatial and spatiotemporally localized topological structures. a** The spatial topology is defined by the zeros of $S(x, y, z)$ and **b** by the zeros of $p(u, v)$, both exhibit similar topological structures in their respective parameter spaces. **c1-e1** The product of a toroidal torus and a set of infinite twisted surfaces [Eq. (7)] yields a topological curve in a spatially nonlocal domain. **c2-e2** The product of a toroidal torus and a set of infinite twisted surfaces [Eq. (2)] yields a topological curve in a localized domain in space and time.

### Acknowledgements

This work was supported by the National Key Research and Development Program of China (2022YFA1404800 [Y.C.]), the National Natural Science Foundation of China (12534014 [Y.C.], 12192254 [Y.C.], W2441005 [Y.C.], 12434012 [Q.Z.]), the Natural Science Foundation of Shandong Province (ZR2025ZD21 [Y.C.], ZR2024QA216 [J.L.], ZR2023YQ006 [C.L.])), and the Taishan Scholars Program of Shandong Province (tsqn202312163 [C.L.]). Q.Z. also acknowledges support by the Key Project of Westlake Institute for Optoelectronics (Grant No. 2023GD007) and Science and Technology Commission of Shanghai Municipality (24JD1402600). S.A.P. acknowledges support from the Natural Sciences and Engineering Research Council of Canada (RGPIN-2025-04064).

## Author contributions

X.L conceived the idea and developed the theory. Y.Z, A.Z and S.A.P verified the theoretical framework. Y.Z and N.Z performed the simulations and experiments. Y.Z, Z.Z, J.L, C.L, and X.L completed the data analysis and visualizations. Y.Z, S.A.P, Q.Z and X.L contributed to drafting and finalizing the manuscript. C.L, S.A.P, Q.Z, Y.C and X.L co-supervised the project. All authors participated in the discussion and review of the manuscript.

## Data availability

The data that support the findings of this study are available from the authors upon reasonable request.

## Conflict of interest

The authors declare no competing interests.

## Supplementary information

See Supplementary information for supporting content.